# Anisotropy of the Seebeck and Nernst coefficients in parent compounds of the iron-based superconductors

Marcin Matusiak[1,*], Michał Babij[1], Thomas Wolf[2]

1. Institute of Low Temperature and Structure Research, Polish Academy of Sciences, ul. Okolna 2, 50-422 Wroclaw, Poland

2. Institute of Solid State Physics (IFP), Karlsruhe Institute of Technology, D-76021, Karlsruhe, Germany

In-plane longitudinal and transverse thermoelectric phenomena in two parent compounds of iron-based superconductors are studied. Namely, the Seebeck ($S$) and Nernst ($v$) coefficients were measured in the temperature range 10 – 300 K for $BaFe_2As_2$ and $CaFe_2As_2$ single crystals that were detwinned *in-situ*. The thermoelectric response shows sizeable anisotropy in the spin density wave state (SDW) for both compounds, while some dissimilarities in the vicinity of the SDW transition can be attributed to the different nature of the phase change in $BaFe_2As_2$ and $CaFe_2As_2$. Temperature dependences of $S$ and $v$ can be described within a two-band model that contains a contribution from highly mobile, probably Dirac, electrons. The Dirac band seems to be rather isotropic, whereas most of the anisotropy in the transport phenomena could be attributed to "regular" hole-like charge carriers. We also observe that the off-diagonal element of the Peltier tensor $\alpha_{xy}$ is not the same for the *a* and *b* orthorhombic axes, which indicates that the widely used Mott formula is not applicable to the SDW state of iron-based superconductors.

**Introduction**

In solid state physics nematicity is understood as a tendency of an electronic system to organize itself in a way that breaks rotational symmetry. This is caused by correlations rather than the anisotropy of the underlying crystal lattice [1]. Evidence for a nematic phase is found in both copper- [2, 3] as well as iron-based [4, 5] superconductors with some suggestions of an intimate relation between the nematicity and superconductivity [6]. This relation is particularly interesting because of a possible enhancement of superconductivity by nematic fluctuations [7, 8].

In this Rapid Communication we report anisotropic behaviour of the thermoelectric response in parent compounds of the iron-based superconductors above and below their magnetic/structural transitions at $T_{tr}$. The thermal gradient was applied in turn along the *a* or *b* orthorhombic axes of the $BaFe_2As_2$ and $CaFe_2As_2$ single crystals. They were detwinned by cooling below $T_{tr}$ under uniaxial pressure and the transport coefficients determined for these two configurations are clearly distinct. However, we do not see a nematic behaviour above $T_{tr}$ in $CaFe_2As_2$. Changes below $T_{tr}$ are consistent with a multiband picture where one of the bands is characterized by the Dirac-fermion energy spectrum. Such a Dirac cone was postulated to occur in the spin density wave (SDW) state of iron based superconductors [9] and it was recently confirmed experimentally for $BaFe_2As_2$ and $SrFe_2As_2$ [10]. Moreover, our results indicate that the conduction bands are not equally anisotropic which has some implications for possible origins of this anisotropy. Another conclusion stemming from the asymmetric magneto-thermoelectric response is the violation of the Mott formula commonly used to interpret thermoelectric data.

**Experimental**

The high-quality $BaFe_2As_2$ single crystals were grown using a self-flux method [11], whereas $CaFe_2As_2$ was grown from Sn flux [12]. For the experiment, square shaped samples were cut out from as grown plate-like single crystals with edges rotated by 45 degree in relation to the tetragonal axes. The sides of the square were about 2 – 2.5 mm and its thickness 0.1 – 0.3 mm.

During the experiment a sample, which was mounted between two clamps made of phosphor bronze, was subjected to a uniaxial pressure by a beryllium copper spring controlled with a stepper motor. For the resistivity ($\rho$) measurements, the electrical contacts were placed at the corners of a sample and the orientations of the voltage and current leads were switched repetitively during the experiment. This allowed the electrical resistivities $\rho_a$ and $\rho_b$ to be determined using the Montgomery method [13]. The Seebeck ($S$) and Nernst ($v$) coefficients along and across the strain direction were measured in two separate runs. The temperature difference along a sample was determined using two Cernox thermometers and a constantan – chromel thermocouple pre-calibrated in magnetic field attached to the sample through 100 µm thick silver leads. Signal leads were made up of long pairs of 25 µm phosphor bronze wires.

**Results**

Changes in the temperature dependences of the electrical resistivity along *a* (long) and *b* (short) orthorhombic axes resulting from the increasing uniaxial pressure along *b* are in good agreement with previously reported data [14, 15]. The anisotropy of $\rho$ rises with the stress and the degree of twin ordering can be judged by the size of the anomaly in the resistivity at the structural/magnetic transition at $T_{tr}$. The uniaxial pressure was increased step-by-step and measurements of the resistivity were repeated until a saturation of the anomaly, indicating maximal detwinning, was achieved. The main difference between $BaFe_2As_2$ and $CaFe_2As_2$, as seen in Fig. 1, is a nematic behaviour above $T_{tr}$ that is present only in the first compound and absent in the latter. This is likely related to the type of the transition which seems to be second order in $BaFe_2As_2$ [16] and first order in $CaFe_2As_2$ [12, 17]. Accordingly, the anisotropy of the resistivity is only detectable above $T_{tr}$ in $BaFe_2As_2$ (up to about 50 K above $T_{tr}$), similarly to fluctuations of the SDW phase in $RFeAsO_{1-x}F_x$ [18]), whereas the electronic system in the tetragonal phase of $CaFe_2As_2$ shows no sign of broken rotational symmetry. Another difference between the two parent compounds studied also manifests itself below $T_{tr}$. Namely, figure 1 shows the variation of the resistivity when a sample is cooled down under uniaxial pressure, which is subsequently released at $T = 3$ K and measurements are continued on warming for a stress-free crystal. While changes of $\rho$ between the down and up ramps in $CaFe_2As_2$ are minimal, $BaFe_2As_2$ clearly goes back to the twinned state deeply in the orthorhombic phase.

The Seebeck coefficient presented in Fig. 2 shows a substantial anomaly in the SDW phase for both parent compounds. In either case, this anomaly is more pronounced along the *b* axis, but both $S_a$ and $S_b$ seem to share the same characteristic features. As expected, previously reported results for the thermopower in a twinned samples lie between $S_a$ and $S_b$ measured here [12, 19] with exemption of the low temperature minimum in $BaFe_2As_2$ that turns out to be somewhat deeper in the present work. In order to verify this we repeated measurements for another $BaFe_2As_2$ crystal (marked with open points in Fig. 2) and the results obtained were basically the same. A probable reason for small discrepancies is the multiband nature of 122 iron pnictides [9], where samples of different origin may exhibit different scattering ratios and different inter-band compensations. Nevertheless, the thermoelectric power measured along *a* axis differs significantly from one measured along *b* axis. And as previously suggested [20] this anisotropy in the SDW phase is much more pronounced than that one observed for the resistivity. To account for small differences between separate measurements of $S_a$ and $S_b$, these two were matched in the high temperature region (270 – 300 K) by multiplication with a suitable correction factor in the region 0.9 – 1.1. Figure 3 compares the normalized anisotropy of the thermoelectric power $\Delta S = (S_b - S_a)/(S_{max} - S_{min})$ for both samples. $\Delta S$ in BaFe$_2$As$_2$ is similar to that reported for EuFe$_2$(As$_{1-x}$P$_x$)$_2$ [20], i.e. it is slightly negative just above $T_{tr}$ and rapidly turns positive just below the transition. However, $\Delta S$ in BaFe$_2$As$_2$ clearly changes sign back to negative at low temperatures, which was not the case of EuFe$_2$(As$_{1-x}$P$_x$)$_2$. This also does not happen in CaFe$_2$As$_2$, where $\Delta S$ stays positive in the entire SDW phase and, in agreement with the resistivity data, there is no sign of anisotropy above $T_{tr}$.

The Nernst coefficient presented in Fig. 4 also shows as a sizeable anisotropy, with $v_a$ and $v_b$ exhibiting a large anomaly below $T_{tr}$. The development of such a maximum in $v(T)$ in the SDW phase of iron pnictides was already reported for CaFe$_2$As$_2$ [12], EuFe$_2$As$_2$ [21] and was attributed to the possible influence of highly mobile Dirac fermions [22]. It is worth noting that the Nernst coefficient at the maximum is about an order of magnitude larger in BaFe$_2$As$_2$ than in CaFe$_2$As$_2$. Measurements of the Nernst effect were carried out for two BaFe$_2$As$_2$ crystals and both sets of data are much alike, as can be seen in Fig. 4 (results obtained for the first sample (solid points) were multiplied by the factor of ~0.85 to account for errors in determining geometrical factors). As was done for the thermopower, $v_a$ and $v_b$ were matched in the high temperature region by applying a small correction. The normalised anisotropy of the Nernst coefficient $\Delta v = (v_b - v_a)/(v_{max} - v_{min})$ presented in Fig. 5, is substantial in the SDW phase. $\Delta v(T)$ in BaFe$_2$As$_2$ is positive ($v_b > v_a$) and roughly proportional to the average Nernst signal, whereas $\Delta v$ in CaFe$_2$As$_2$ is negative just below $T_{tr}$, then $\Delta v$ becomes positive at low temperature i.e. $v_b > v_a$. Analogously to the resistivity and thermopower, there is no anisotropy of the Nernst effect in the tetragonal phase of CaFe$_2$As$_2$, whereas in BaFe$_2$As$_2$ a contribution from fluctuations appears below ~200 K.

**Discussion**

There is a kind of incongruity between the Nernst and resistivity results. Namely, at low temperature both quantities are larger along *b* axis, i.e. $v_b > v_a$ as well as $\rho_b > \rho_a$ (thus the electrical conductivities $\sigma_b < \sigma_a$). This might seem contradictory, since within the semi-classical approximation $\nu = \frac{\pi^2 k_B}{3e} \frac{k_B T}{\epsilon_F} \mu$ [23], where $k_B$ is the Boltzmann constant, $e$ is the electronic charge, $\epsilon_F$ is the Fermi energy and $\mu$ is the mobility and since $\sigma = ne\mu$ ($n$ is the charge carrier concentration), one can expect a proportional relation between $\sigma$ and $\nu$ for a single band. Nevertheless, the situation can be different in a multiband case and the iron-arsenides are known to be multiband conductors [24]. Furthermore, band calculations predict that one of the conducting bands in the SDW phase is a topologically protected Dirac cone [9], whose presence in the 122 iron pnictides was repeatedly confirmed by angle-resolved photoemission spectroscopy [25], quantum oscillations [26], and recently by infrared studies [10]. Temperature dependences of the transport coefficients [12, 21] are in good agreement with calculations performed within a simplified model consisting of a hole band (denoted below with index $h$) with a conventional energy spectrum, and an electron band (index $e$) with the Dirac-fermion energy spectrum [22]. The total Nernst coefficient is the sum of contributions from different bands weighted by the respective electrical conductivities (under assumption that an ambipolar term is small) and $\sigma = \sigma^e + \sigma^h$:

$$\nu = \frac{\sigma^e \nu^e + \sigma^h \nu^h}{\sigma^e + \sigma^h}. \tag{1}$$

Because of a high mobility of Dirac fermions, which was observed for example in EuFe$_2$As$_2$ [21], one can expect that $\nu^e \gg \nu^h$. In such a case there are three possibilities:

i) If in-plane anisotropy of both bands are comparable then decreasing $\sigma^e$ and $\sigma^h$ along with decreasing $\nu^e$ and $\nu^h$ will cause total $\nu$ along $b$ axis to be smaller than $\nu$ along $a$ axis ($\nu_b < \nu_a$), as in a single band scenario;

ii) If anisotropy of the Dirac (electron-like) band is more pronounced then again one will observe $\nu_b < \nu_a$, since the numerator of Eqn. 1 will be dominated by the varying $\sigma^e \nu^e$ term;

iii) In contrast, if the Dirac band is isotropic, whereas anisotropy can be attributed to the hole-like band than for $\nu^e \gg \nu^h$ (and independently of the signs of $\nu^e$ and $\nu^h$) the denominator in Eqn. 1 will decrease faster than the numerator leading to the "reverse" relation, i.e. $\sigma_b < \sigma_a$ and $\nu_b > \nu_a$.

Therefore, our results not only indicate a presence of at least two bands in the SDW phase of 122 compounds, but primarily suggest that the Dirac band is rather isotropic, whereas anisotropy should be attributed to the regular hole-like carriers. Subsequently, since the Dirac band seems to be very sensitive to a shift of the chemical potential [21], this would point towards anisotropic scattering [27] rather than orbital polarization [28] as a main reason for the anisotropy below $T_{tr}$.

Here we would like to focus on another important conclusion having general repercussions for analysis of the thermoelectric data in the iron-based superconductors. We start with realisation that the Nernst coefficient is in fact composed of two contributions. This is sometimes called the Sondheimer cancellation, that in an anisotropic material takes a form: $\nu B = -\alpha_{yx}/\sigma_{yy} - S\,\sigma_{xy}/\sigma_{yy}$ ($\alpha_{yx}$ is the off-diagonal element of the Peltier tensor, $\sigma_{xy}$ is the Hall conductivity), as long as $\sigma_{xx} \neq \sigma_{yy}$, $\sigma_{yy} \gg \sigma_{xy}$ and $\sigma_{xy} = -\sigma_{yx}$. The last condition results from the Onsager reciprocal relation $\sigma_{xy}(B) = \sigma_{yx}(-B)$ in a system obeying the reflection symmetry (which applies also to the SDW phase of the iron-pnictides). The $S\,\sigma_{xy}/\sigma_{yy}$ term can be calculated using previously reported data for the Hall coefficient [12, 29] and its temperature dependences are presented in Fig. 4. Consequently, these data allow one to determine $\alpha_{yx}$ for *a* and *b* orientations and the results are shown in Fig. 6. For both compounds there are clear differences between $\alpha_{yx}$ determined for $\Delta T \parallel a$ and for $\Delta T \parallel b$, which in inset in Fig. 6 were normalised analogously to the Seebeck and Nernst anisotropy, i.e. $\Delta\alpha_{yx}^{norm} = 1/B\,(\alpha_{yx}^b - \alpha_{yx}^a)/(\alpha_{yx}^{max} - \alpha_{yx}^{min})$. Evidently in either case $\Delta\alpha_{yx}^{norm}$ increases significantly at low temperature, which means that $\alpha_{xy} \neq -\alpha_{yx}$. This in turn indicates that the generalised Mott relation $\alpha_{ij} = -\frac{\pi^2 k_B}{3e}\left(\frac{\partial \sigma_{ij}}{\partial \epsilon}\right)_{\epsilon_F}$ [30], which is commonly used to interpret thermoelectric data, fails. There may be several potential reasons for such a violation, since the Mott formula is strictly valid only for non-interacting particles. First thing to note is that the nematic order itself can be a manifestation of strong electronic correlations [31, 32]. On the other hand, the ground state of iron-pnictides seems to be well described as a Fermi liquid, showing conventional quantum oscillations [26] and obeying the Wiedemann-Franz law in the low and high temperature limits [33]. Another possibility is related to the very small Fermi energy in 122 compounds [21, 29, 34] where the Fermi temperature ($T_F$) can be as low as about 100 K, since the Mott formula relies on a Sommerfeld expansion and is only valid for $T \ll T_F$ [35]. Such a scenario might be supported by the tendency that $\alpha_{xy} = -\alpha_{yx}$ in the low temperature limit for CaFe$_2$As$_2$. This is not observed in BaFe$_2$As$_2$ perhaps because the estimate $T_F \approx 115$ K for BaFe$_2$As$_2$ [29] is about two times smaller than $T_F$ for CaFe$_2$As$_2$ ($T_F \approx 230$ K [12]). The very large value of the Nernst coefficient below $T_{tr}$ in BaFe$_2$As$_2$ also suggests extremely small $T_F$ since, as mentioned above, $\nu$ should be roughly proportional to $1/\epsilon_F$ [23]. However, the decrease of $\Delta\alpha_{yx}^{norm}$ with increasing temperature, which goes to zero in the tetragonal state without any obvious anomaly, suggests that the reason for violation of the Mott formula might be different. A likely candidate is the electron-electron (e-e), or more generally inelastic, scattering that can cause deviations from the Mott formula [36] and such e-e processes were suggested to play an important role in iron-pnictides [37].

**Summary**

We studied the in-plane anisotropy of magneto-thermoelectric phenomena in single crystals of the parent compounds of iron-based superconductors. For both compositions the Seebeck and Nernst coefficients were measured along the *a* and *b* crystallographic axes and the anisotropies observed are much larger than those in the electrical resistivity. However, there are some differences in the thermoelectric response between $BaFe_2As_2$ and $CaFe_2As_2$, perhaps because there are SDW fluctuations only for the first compound. The results are consistent with the scenario in which electronic transport in the SDW state consists of two contributions – one from the electron-like Dirac cone and another one from a regular hole-like band. We are able to conclude that the first, highly mobile one, is rather isotropic, whereas the hole-like band is mostly responsible for the observed anisotropy. This also suggests that the nematicity observed in 122 iron-based superconductors may be related to scattering rather than band polarisation. Another important observation, that $\alpha_{xy} \neq -\alpha_{yx}$, indicates that the Mott relation fails in the SDW phase of the iron-pnictides.

**Acknowledgments**

The authors would like to thank D. Orgad and V. Oganesyan for helpful comments.
This work was supported financially by the National Science Centre (Poland) under the research Grant No. 2014/15/B/ST3/00357.

**Figures**

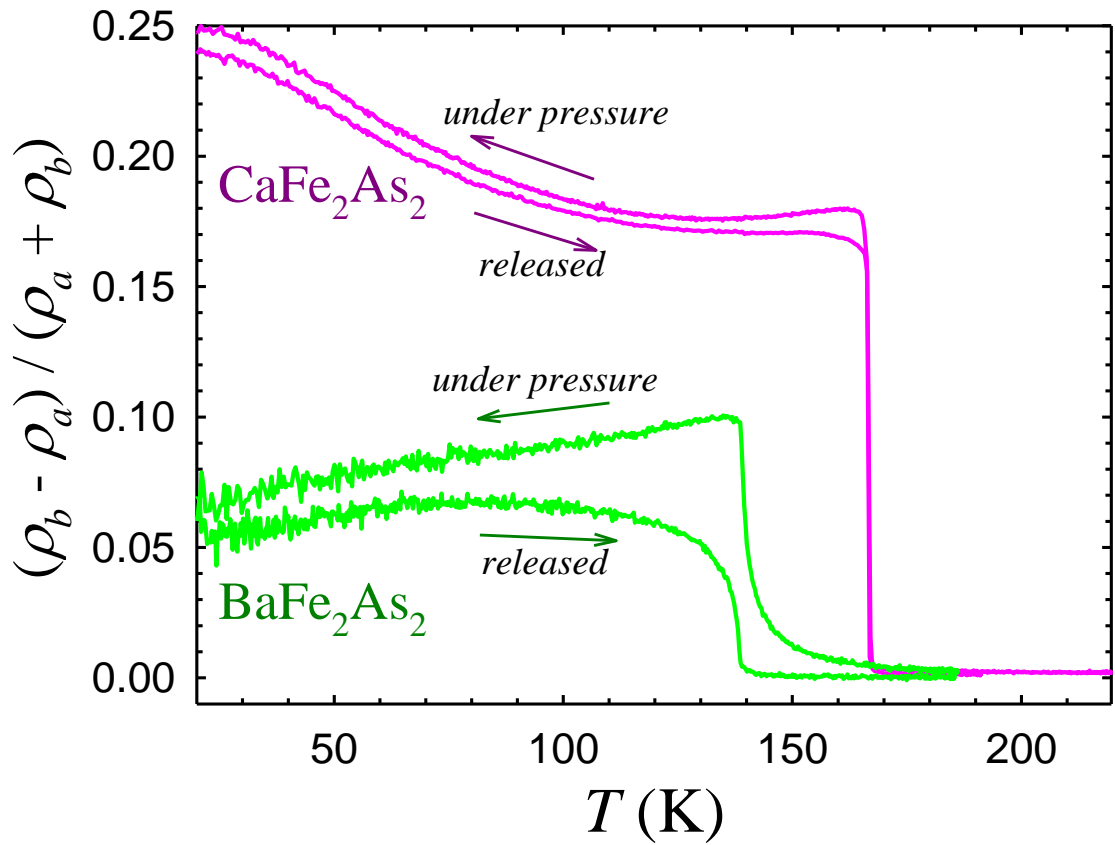

**Figure 1.**
(Color online) The temperature dependences of the normalised resistivity anisotropy when a sample is cooled under pressure which is released at low temperature and afterwards another set of data is collected on warming.

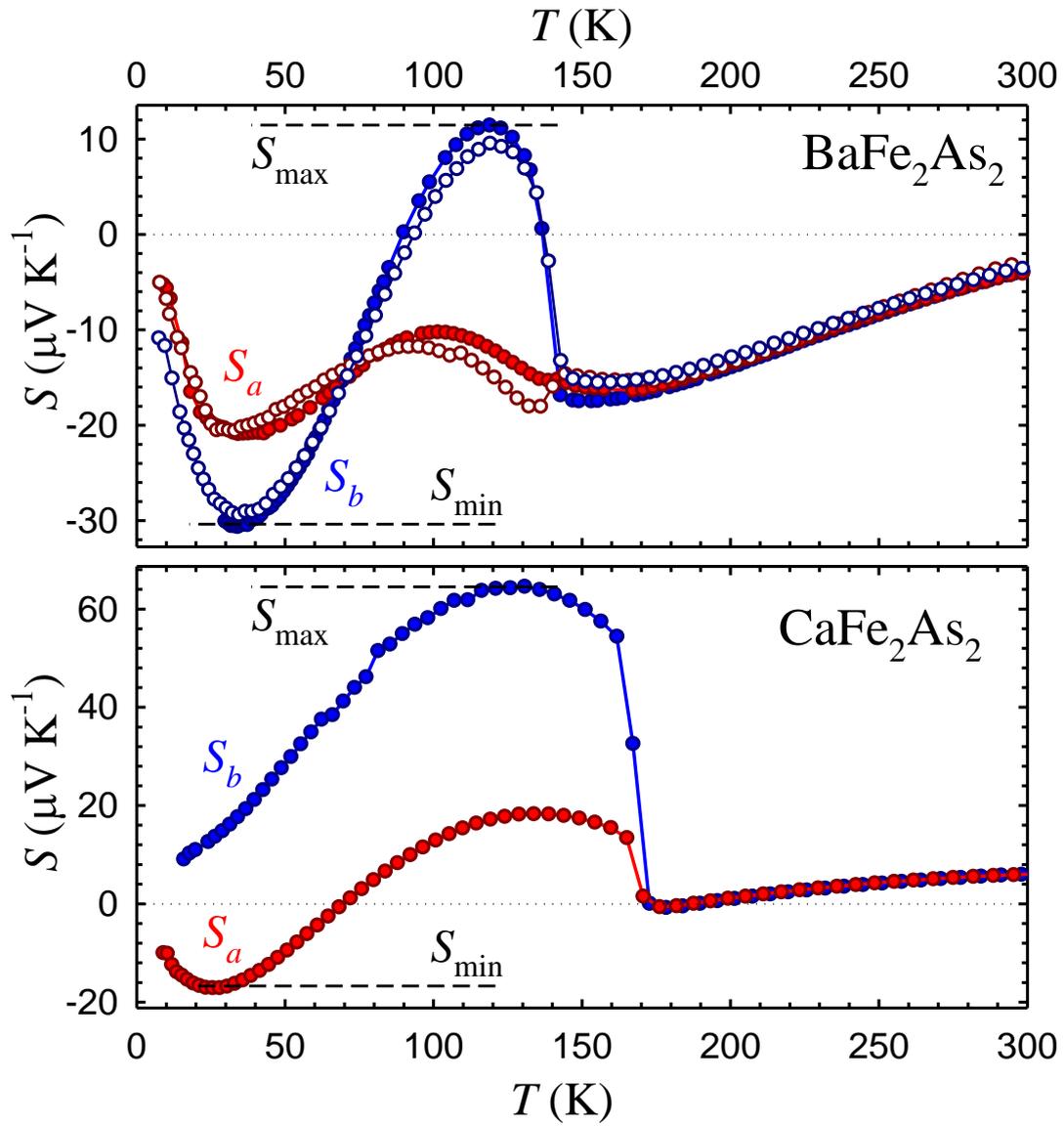

**Figure 2.**
(Color online) The temperature dependences of the thermoelectric power in BaFe$_2$As$_2$ (solid points: sample #1, open points: #2) and CaFe$_2$As$_2$ along the orthorhombic *a* and *b* axes.

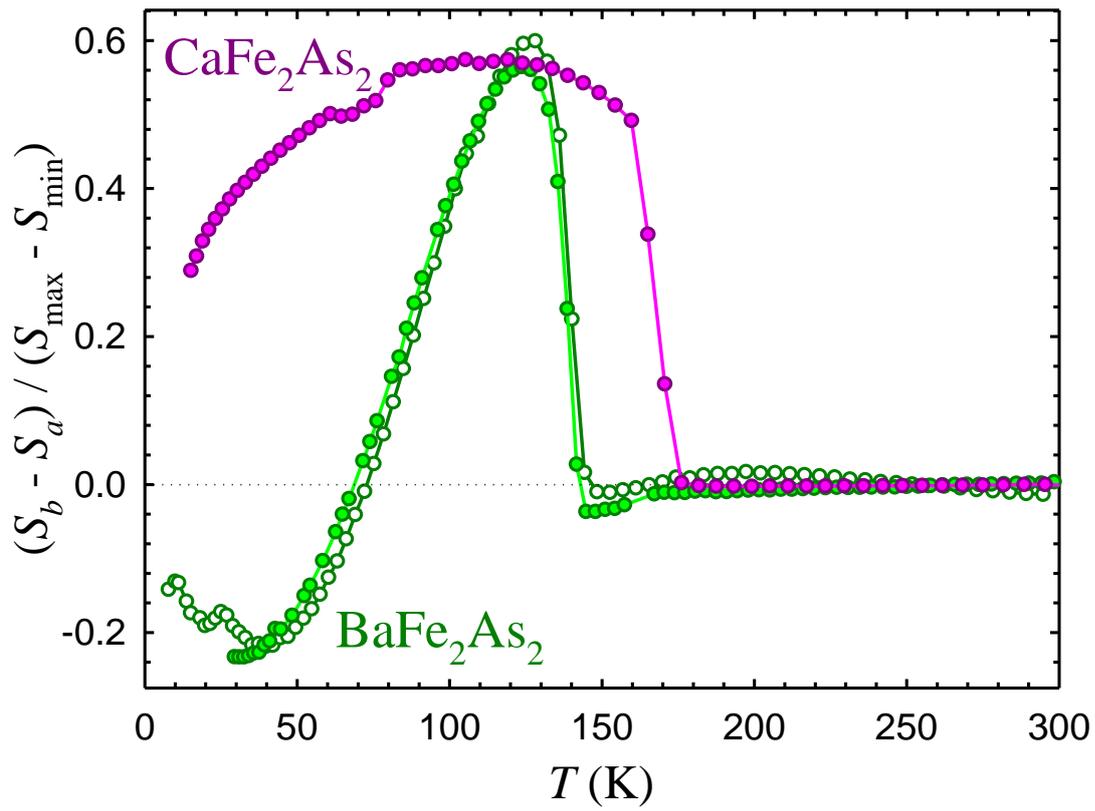

**Figure 3.**
(Color online) The temperature dependences of the normalised thermopower anisotropy in BaFe$_2$As$_2$ (solid points: sample #1, open points: #2) and CaFe$_2$As$_2$.

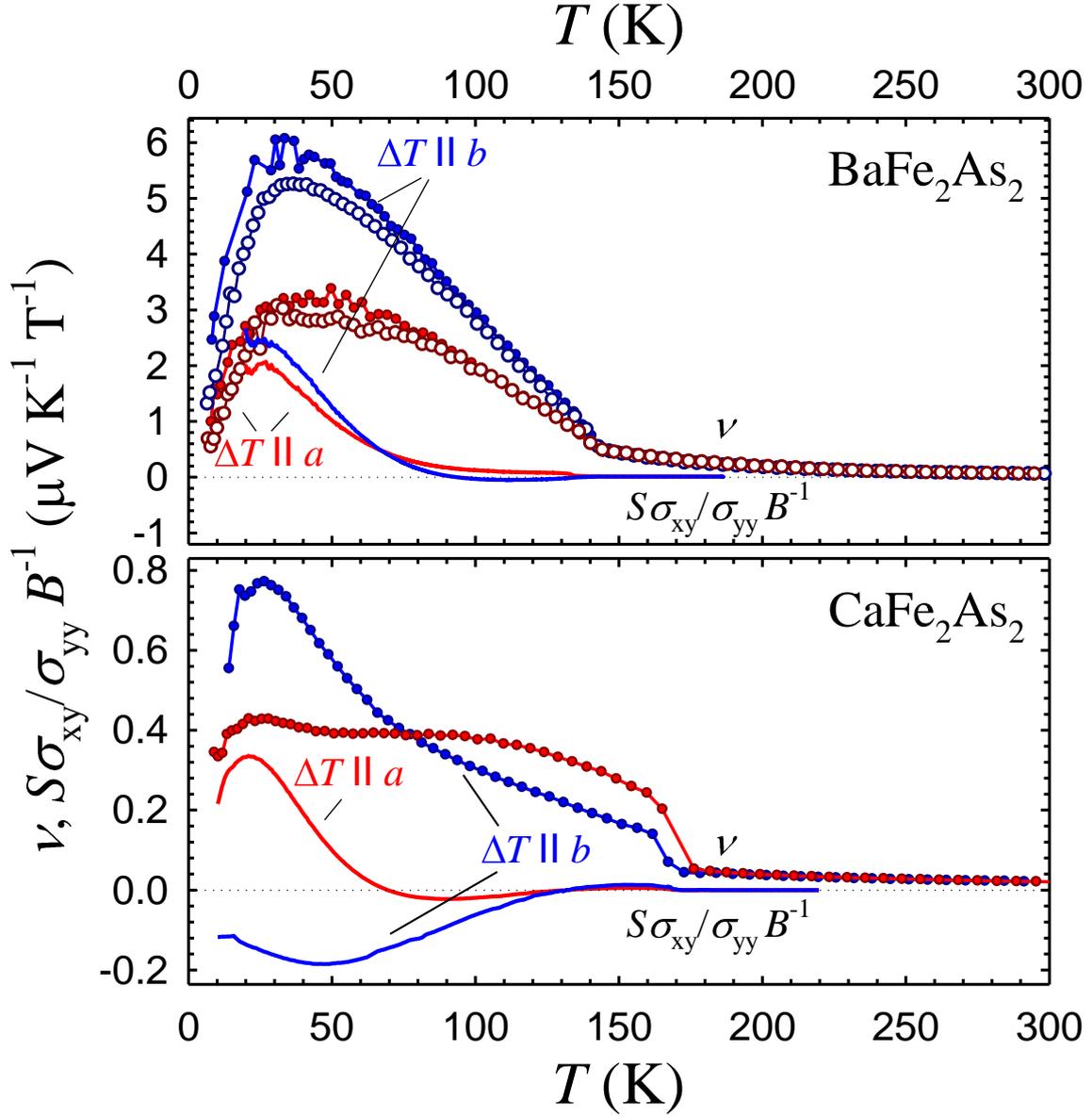

**Figure 4.**
(Color online) The temperature dependences of the Nernst coefficient (points) and the $S\sigma_{xy}/\sigma_{yy}B^{-1}$ term (solid lines) along the orthorhombic $a$ and $b$ axes for BaFe$_2$As$_2$ (solid points: sample #1, open points: #2) and CaFe$_2$As$_2$.

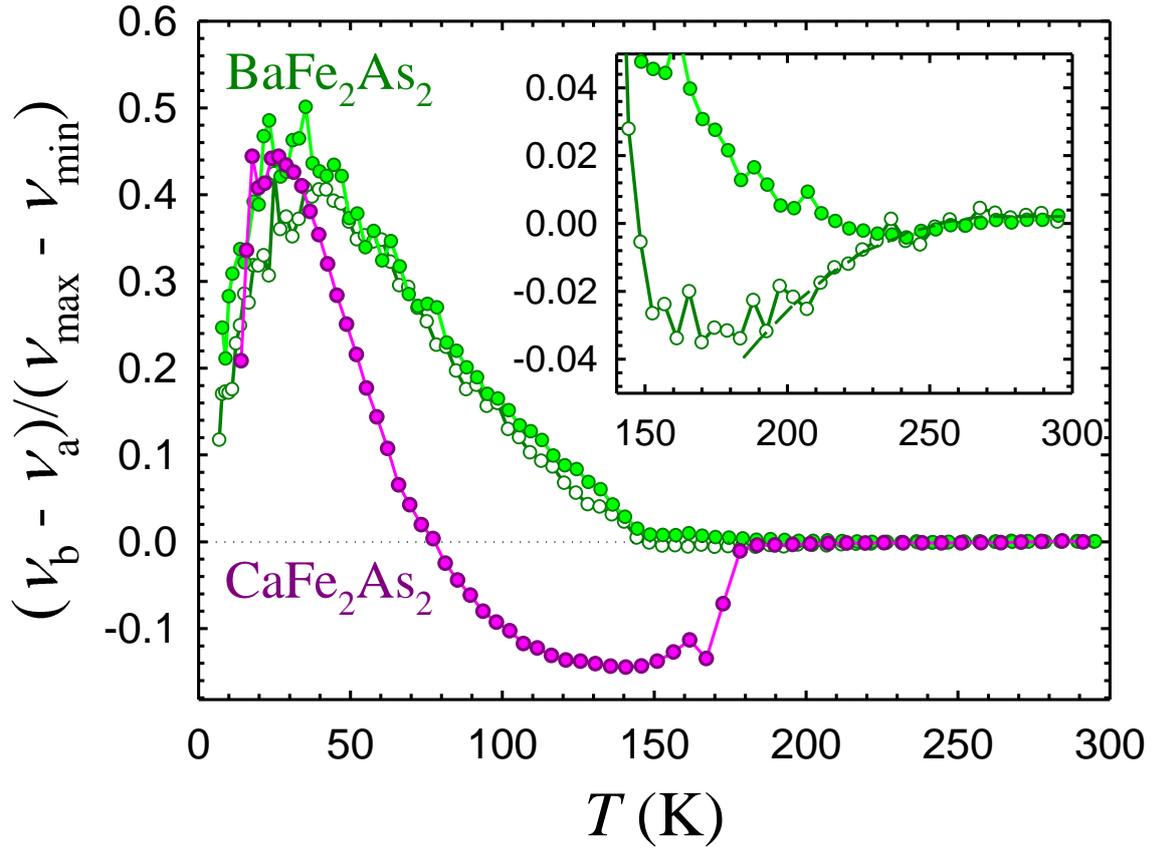

**Figure 5.**

(Color online) The temperature dependences of the normalised Nernst effect anisotropy in BaFe$_2$As$_2$ (solid points: sample #1, open points: #2) and CaFe$_2$As$_2$. Inset shows the high temperature BaFe$_2$As$_2$ data in enlarged scale to highlight an onset of the anisotropy.

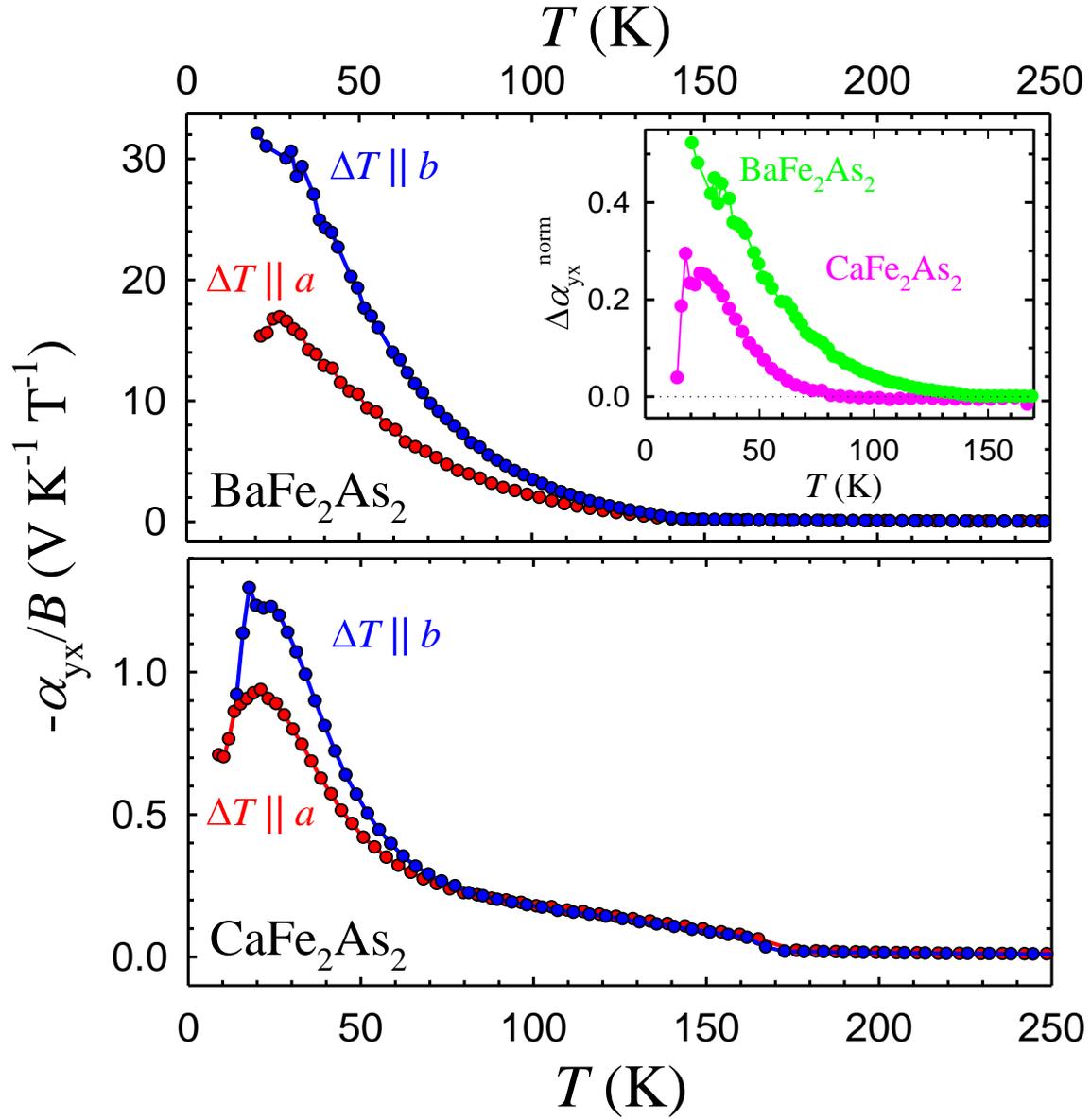

**Figure 6.**
(Color online) The temperature dependences of the off-diagonal Peltier element $-\alpha_{yx}$ along the orthorhombic *a* and *b* axes. Inset shows the normalised anisotropy of $\alpha_{yx}$ for $BaFe_2As_2$ and $CaFe_2As_2$.